\documentclass[twocolumn,showpacs,preprintnumbers,amsmath,amssymb]{revtex4}

\usepackage{epsf}

\begin{document}

\title{Estimating the distribution of dynamic invariants: illustrated
with an application to human photo-plethysmographic time series}
\author{Michael Small\thanks{Tel: +852 2766 4744 , 
Fax: +852 2362 8439, email: {\tt ensmall@polyu.edu.hk}.}}
\affiliation{Department of Electronic and Information
Engineering\\ Hong Kong Polytechnic University, Hung Hom, Kowloon,
Hong Kong} 

\date{\today} 

\begin{abstract}
Dynamic invariants are often estimated from experimental time series
with the aim of differentiating between different physical states in the
underlying system. The most popular schemes for estimating dynamic
invariants are capable of estimating confidence intervals, however such
confidence intervals do not reflect variability in the underlying
dynamics. In this communication we propose a surrogate based method to
estimate the expected distribution of values under the null hypothesis
that the underlying deterministic dynamics are stationary. We
demonstrate the application of this method by considering four
recordings of human pulse waveforms in differing physiological states
and provide conclusive evidence that correlation dimension is capable of
differentiating between three (but not all four) of these states.
\end{abstract}


\maketitle

Various dynamic invariants are often estimated from time series in a
wide variety of scientific disciplines. It has long been known that
these estimates (and in particular correlation dimension
estimates) alone are not sufficient to differentiate between chaos and noise. Most
notably, the method of surrogate data \cite{jT92} was introduced in an
attempt to reduce the rate of false positives during the hunt for physical
examples of chaotic dynamics. Although it is not possible to find
conclusive evidence of chaos through estimation of dynamic invariants,
surrogate methods are used to generate a distribution of statistic
(i.e. the estimates of the dynamic invariant) values under the
hypothesis of linear noise. In the most general form, the standard
surrogate methods can generate the distribution of statistic values under
the null hypothesis of a static monotonic nonlinear transformation of
linearly filtered noise.

In this communication, we introduce a significant generalisation of a recent
surrogate generation algorithm \cite{cyclsurr,pps2}. The {\em pseudo-periodic
surrogate}  (PPS) algorithm allows one to generate data consistent with
the null hypothesis of a noise driven periodic orbit --- provided the
data exhibits pseudo-periodic dynamics. This algorithm has been
applied to differentiate between a noisy limit cycle, and deterministic
chaos. By modifying this algorithm and applying it to noisy time series
data, we are able to generate surrogate time series that are independent
trajectories of the same deterministic system.

This ensemble of {\em attractor trajectory surrogates} (ATS) can then be used
to estimate the distribution statistic values for estimates of
any statistic derived from these time series. The statistics of greatest
interest
to us are dynamic invariants of the underlying attractor, and in
particular correlation dimension and entropy estimates provided by the
{\em Gaussian kernel algorithm} (GKA) \cite{cD96,effgka}. Our choice of the
GKA is entirely arbitrary, but based on our familiarity with this
particular algorithm.

An important application for the ATS technique is to determine whether
dynamic invariants estimated from distinct time series are significantly
different. The question this technique can address is whether (for
example) a correlation dimension of 2.3 measured during normal
electrocardiogram activity is really distinct from the correlation
dimension of 2.4 measured during an episode of ventricular tachycardia
\cite{csf,cic4}. Estimates of dynamic invariants (including the GKA
\cite{cD96,effgka}) often come with confidence intervals. But these
confidence intervals can only be based on uncertainty in the
least-mean-square fit, not the underlying dynamics. Conversely, it is
standard practice to obtain a large number of representative time series
for each (supposedly distinct) physical state, and compare the
distribution of statistic values derived from these. But, this approach
is not always feasible: in \cite{csf,cic4} for example, the problem is not
merely that these physiological states are both difficult and dangerous to
replicate, but that inter-patient variability makes doing so infeasible.

In the remainder of this communication we describe the new ATS algorithm
and demonstrate that it can be used to estimate the distribution of
dynamic invariant estimates from a single time series of a known
dynamical system (the chaotic R\"ossler system). We then apply this same
method to four recordings of human pulse waveforms, measured via
photo-plethysmography \cite{jB99,jB01}. Each of the four recordings
correspond to a distinct physiological state. We compute correlation
dimension and entropy using the GKA method and show that the expected
distribution of correlation dimension and entropy estimates are
sufficient to differentiate between these four physiological states.

The ATS algorithm may be described as follows. Embed a scalar time
series $\{x_t\}$ to obtain a vector timeseries $\{z_t\}$ (of length
$N$). The choice of embedding is arbitrary, but has been adequately
discussed in the literature (\cite{window} for example). From the
embedded time series, the surrogate is obtained as follows. Choose an
initial condition, $w_1\in\{z_t|t=1,\ldots,N\}$. Then, at each step,
choose the successor to $w_t$ with probability
\begin{eqnarray}
\label{switch}
P(w_{t+1}=z_{i+1}) & \propto &
\exp{\frac{-\|w_t-z_i\|}{\rho}}
\end{eqnarray}
where the {\em noise radius} $\rho$ is an as-yet unspecified
constant. In other words, the successor to $w_t$ is the successor of a
randomly chosen neighbour of $w_t$. Finally, from the vector time series
$\{w_t\}$ the ATS $\{s_t\}$ is obtained by projecting $w_t$ onto
$[1\;0\;0\;0\;\cdots\;0]$ (the first coordinate). Hence
\begin{eqnarray}
s_t & = & w_t\cdot[1\;0\;0\;0\;\cdots\;0]
\end{eqnarray}

In \cite{cyclsurr,pps2} this algorithm was shown to be capable of
differentiating between deterministic chaos and a noisy periodic
orbit. In the context of the current communication we assume that
$\{x_t\}$ is contaminated by additive (but possibly dynamic) noise and
we choose the noise radius $\rho$ such that the observed noise
is replaced by an independent realisation of the same noise
process. Furthermore, we assume
that the deterministic dynamics are preserved by suitable choice of
embedding parameters. Under these two assumptions, $\{z_t\}$ and
$\{w_t\}$ have the same invariant density and $\{x_t\}$ and $\{s_t\}$
are therefore (noisy) realisation of the same dynamical system with (for
suitable choice of $\rho$) the same noise distribution.

As in \cite{cyclsurr,pps2} the problem remains the correct choice of
$\rho$. This is the major difference between the ATS described here and the PPS of
\cite{cyclsurr,pps2}. However, since the null hypothesis we wish to
address is different from (and more general than) that of the PPS,
choice of $\rho$ for the ATS is less restrictive. For $t=T$ given, one
can compute $P(w_{t+1}\neq z_{i+1}
\wedge \|w_t-z_i\|=0 | t=T)$ directly from the data by applying (\ref{switch}). Assuming the
process is ergodic \footnote{This assumption is sufficient rather than
necessary.} one can then sum
\begin{eqnarray}
\lefteqn{P(w_{t+1}\neq z_{i+1} \wedge \|w_t-z_i\|=0) = }\\
\nonumber && \frac{1}{N}\sum_{T=1}^N
P(w_{t+1}\neq z_{i+1} \wedge \|w_t-z_i\|=0 | t=T)
\end{eqnarray}
to get the probability of a temporal discontinuity
\footnote{By temporal discontinuity we mean that $w_t=z_i$ but
$w_{t+1}\neq z_{i+1}$.} in the
surrogate at any time instant. There is a 1:1 correspondence between a
value $p=P(w_{t+1}\neq z_{i+1} \wedge \|w_t-z_i\|=0)$ and $\rho$, and we
choose to implement (\ref{switch}) for a particular value of $p$ (i.e. a
particular transition probability) rather than a specific noise
level. In what follows we find that studying intermediate values of $p$
($p\in[0.05,0.95]$) is sufficient. However, the significant point is
that $p\in[0.05,0.95]$ corresponds to a very narrow range of values of
$\rho$.

\begin{figure}
  \[\epsfxsize 75mm \epsfbox{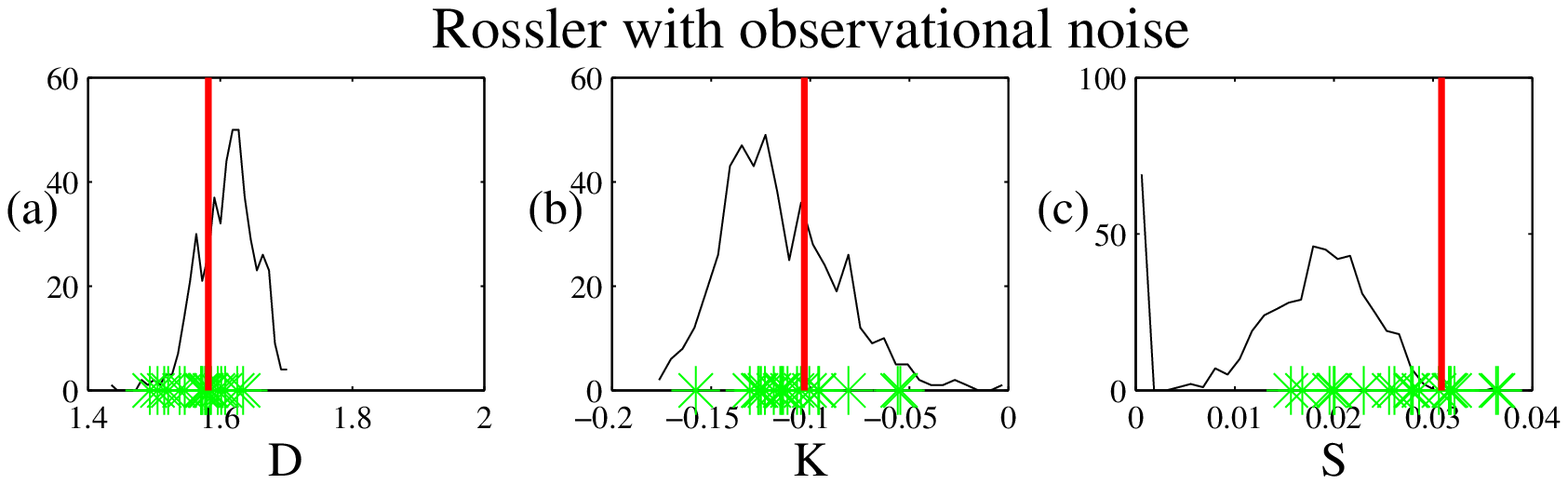}\]
  \[\epsfxsize 75mm \epsfbox{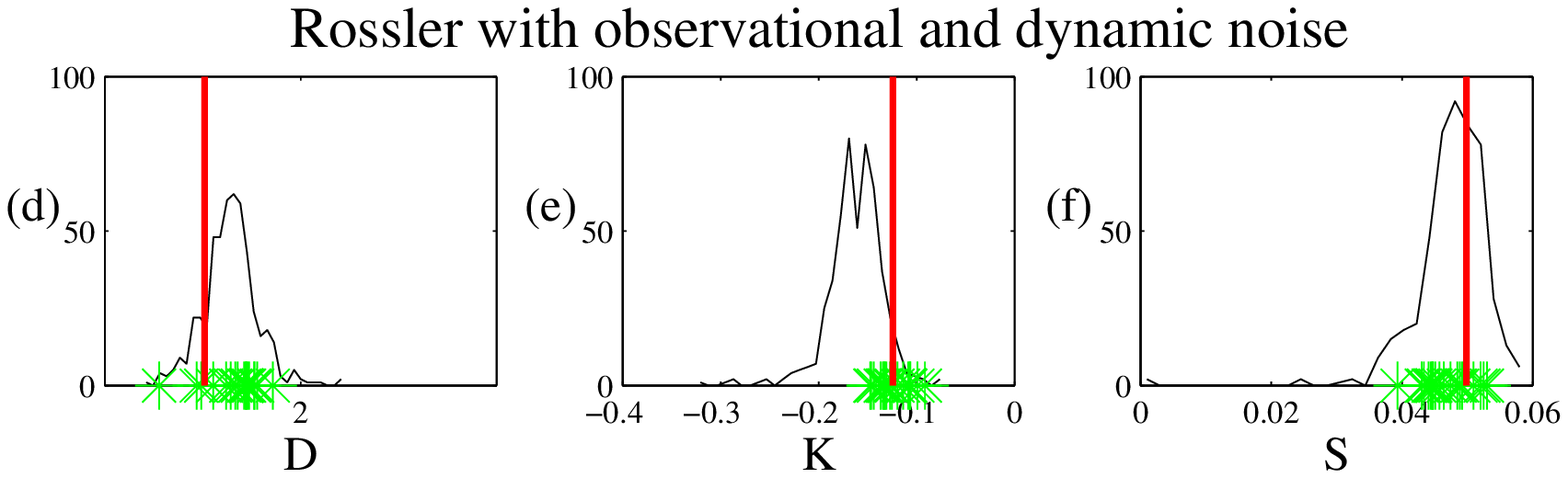}\]
\caption{{\bf Distribution of statistics $D$, $K$ and $S$ for
short and noisy realisations of the R\"ossler system.} The histogram
shows the distribution of statistic estimates ($D$, $K$ and $S$) for
$500$ ATS time series generated from a $1000$ point realisation of the
R\"ossler system. The solid vertical line on each plot is the comparable
value for the data and the stars marked on the horizontal axes are for
$20$ independent realisations of the same process. The top row of
figures depicts results for the R\"ossler system with observational
noise only, the bottom row of figures has both observational and dynamic
noise. Panels (a) and (d) show correlation dimension estimates, (b) and
(e) are entropy, and (c) and (f) are noise level.}
\label{rossler}
\end{figure} 

We now demonstrate the applicability of this method for noisy time
series data simulated from the R\"ossler differential equations (during
``broad-band'' chaos). We integrated ($1000$ points with a time step of
$0.2$) the R\"ossler equations both with and without multidimensional dynamic noise at
$5\%$ of the standard deviation of the data. We then studied the
$x$-component after the addition of $5\%$ observational noise. We
selected embedding parameters using the standard methods ($d_e=3$ and
$\tau=8$) and then compute ATS surrogates for various exchange
probabilities $p=0.05,0.1,0.15,\ldots,0.95$. For the data set and each
ensemble of surrogates we then estimated correlation dimension $D$,
entropy $K$ and noise level $S$ using the GKA algorithm
\cite{cD96,effgka} (GKA embedding using embedding dimension $m=2,3,\ldots,10$ and
embedding lag of $1$). Figure \ref{rossler} depicts the results when the GKA is
applied with embedding dimension $m=4$ and the exchange probability is
$p=0.35$. Other values of $m$ gave equivalent results, as did various
values of $p$ in the range $[0.2,0.8]$.

For $p\in[0.2,0.8]$ we found that the estimate of noise $S$ from the GKA
algorithm coincided for data and surrogates, but this was often not the
case for extreme values of $p$. Therefore, this estimate of signal noise
content is a good indicator of the accuracy of the dynamics reproduced
by the ATS time series. Furthermore to confirm the spread of the data we
also estimated $D$, $K$, and $S$ for $20$ further
realisations of the same R\"ossler system (with different initial
conditions). In each case, as expected, the range of these values lies
well within the range predicted by the ATS scheme.

\begin{figure}
  \[\epsfxsize 75mm \epsfbox{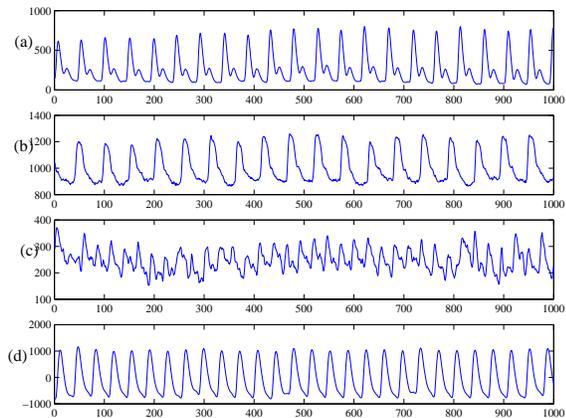}\]
\caption{{\bf Human pulse waveform recorded with photo-plethysmography.}
Four recordings of human pulse waveform (61 Hz) in four different
physiological conditions. The four time series correspond to: (a)
normal, (b) quasi-stable, (c) unstable, and (d) post-operative (stable).}
\label{ppg}
\end{figure} 

We now consider the application of this method to photo-plethysmographic
recordings of human pulse dynamics over a short time period (about 16.3
seconds). We have access to only a limited amount of data representative
of each of four different dynamic regimes. In any case, we would expect
the system dynamics to change if measured over a significantly longer
time frame. The data collection and processing with the methods of
nonlinear time series analysis are described in
\cite{jB99,jB01}. Previously, we have studied nonlinear determinism in
cardiac dynamics measured with electrocardiogram (ECG)
\cite{csf,cic4}. Although we do not consider ECG data here, this data would
be another useful system to examine with these methods
\footnote{Actually, the problem here is that we have too much data and
it is therefore difficult to select a ``representative'' small number of
short time series. However, we intend to examine this data more
carefully in forthcoming work.}. The four data sets we examine in this
communication are depicted in figure \ref{ppg}.

\begin{figure}
  \[\epsfxsize 75mm \epsfbox{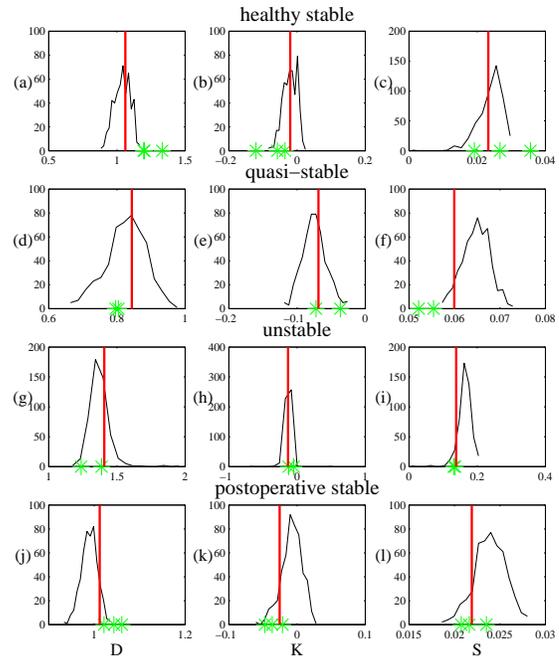}\]
\caption{{\bf Distribution of statistics $D$, $K$ and $S$ for
human pulse waveforms.} The histogram shows the distribution of
statistic estimates ($D$, $K$ and $S$) for $500$ ATS time series
generated from each of the four time series depicted in figure
\ref{ppg}. The solid vertical line on each plot is the comparable value for the data
and the stars marked on the horizontal axes are for the (limited)
subsequent data recorded from each patient. In each case only two or
three subsequent contiguous but non-overlapping timeseries were
available. The figures are: (a) correlation dimension ($D$), (b) entropy
($K$), and (c) noise ($S$) for the normal rhythm; (d) $D$, (e) $K$, and
(f) $S$ for the quasi-stable rhythm; (g) $D$, (h) $K$, and (i) $S$ for
the unstable rhythm; and (j) $D$, (k) $K$, and (l) $S$ for the
post-operative stable rhythm.}
\label{ppgres1}
\end{figure} 

For each data set we repeated the analysis described for the R\"ossler
time series. Results for GKA embedding dimension $m=4$ and $p=0.35$
are depicted in figure \ref{ppgres1}. As with the R\"ossler system,
variation of the parameters $m$ and $p$ did not significantly change
the results. We find that in every case (except for $p\notin[0.2,0.8]$)
the distribution of $D$, $K$ and $S$ estimated from the ATS data using
the GKA included the true value. Most significantly, this indicates that
the range of values of $p$ is appropriate. Moreover, these results are
consistent with the hypotheses that
the noise is effectively additive and can be modelled with this simple
scheme, and that the underlying deterministic dynamics can be
approximated with a local constant modelling scheme. 

We also estimated the statistics $D$, $K$ and $S$ for
additional available data (subsequent, contiguous, but non-overlapping)
from each of the four rhythms. This small amount of data afforded us two
or three additional estimates of each statistic for each rhythm. For the
unstable and quasi-stable rhythm we observed good agreement. For the
stable (normal and post-operative) rhythms, this is not the case. On
examination of the data we find that this result is to be
expected. Both the stable rhythms undergo a change in amplitude and
baseline subsequent to the end of the original $16$ second recording,
this non-stationarity is reflected in the results. This same
non-stationarity has also been observed independently in Bhattacharya
and co-workers \cite{jB99,jB01}.

\begin{figure}
  \[\epsfxsize 85mm \epsfbox{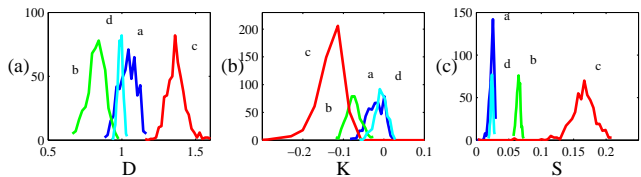}\]
\caption{{\bf Discriminating power of the statistics $D$, $K$ and $S$ for
human pulse waveforms.} The distribution (a binned histogram) of
statistic values estimated via the ATS method (as described in figure
\ref{ppgres1}) for each of the four distinct physiological waveforms is
shown. The four rhythms are labelled 'a', 'b', 'c', and 'd'
corresponding to the same labelling in figure \ref{ppg}.  These figures
show that correlation dimension alone is sufficient to differentiate
between three of these four physiological states. The exception is that
these three statistics are insufficient to differentiate between the
normal and post-operative states.}
\label{ppgres2}
\end{figure} 

We now return to the question that the ATS test was designed to address:
can we differentiate between these four rhythms based on the GKA?
Figure \ref{ppgres2} provides the answer. In figure \ref{ppgres2} we see
the estimated distribution of statistic values ($D$, $K$ and $S$) for
each of the four rhythms shown in figure \ref{ppg}. Clearly (and not
surprisingly), the correlation dimension estimate and noise level of the
unstable rhythm is significantly different from the other three
rhythms. More significantly, the quasi-stable rhythm is also observed to
be distinct from the other three regimes. Furthermore, we observe that
in the quasi-stable state the correlation dimension estimate is
significantly less than one, while for the unstable state it is
significantly larger than one. For example, the quasi-stable state may
be characterised by a noise driven stable focus
\footnote{Due to the discretisation necessary to digitise this data, a
noise drive stable periodic orbit is also a plausible cause of the
observed data. To distinguish these two, a more detailed study is
required.}, while the unstable state exhibits high dimensional
(i.e. $D>1$) deterministic dynamics. Both these regimes exhibit
significantly more (additive Gaussian) noise than the stable states.

The two stable states (panels (a) and (d) of figure \ref{ppg}) are
harder to distinguish: both visually and using the statistics $D$, $K$,
and $S$. While the individual data sets we depict in figure \ref{ppg}
exhibit different statistic values (for example $D=1.06$ and $D=1.01$),
we find that the ATS analysis indicates that these statistics are not
significantly different. Both regimes exhibit a correlation dimension of
about one, and a similar noise level. The variation in observed results
is lesser in the post-operative stable regime, but the distribution do
overlap.

Finally, we find that entropy estimated with the GKA algorithm $K$ is of
no use in differentiating between these four rhythms.

The results of this analysis are in general agreement with those
presented in \cite{jB99,jB01}. Independent linear surrogate analysis
\cite{jT92} has confirmed that each of these four rhythms is inconsistent
with a monotonic nonlinear transformation of linearly filtered noise
\footnote{These calculations are routine, and not presented in this
communication.}. The only significant difference is that the correlation
dimension estimates we present here are significantly lower than those in
\cite{jB99,jB01}. This is due to the different correlation dimension
algorithm. Unlike the algorithm employed in \cite{jB99,jB01}, the GKA
seperates the data into purely deterministic and stochastic components,
and hence estimates both $D$ and $S$. The correlation dimension
estimated in \cite{jB99,jB01} is the combined effect of both components
of the GKA.

Although we have considered the specific application of human pulse
dynamics, the algorithm we have proposed may be applied to a wide
variety of problems. We have shown that provided time delay embedding
parameters can be estimated adequately, and an appropriate value of the
exchange probability is chosen, the ATS algorithm generates independent
trajectories from the same dynamical system. When applied to data from
the R\"ossler system we confirm this result, and we demonstrate its
application to experimental data. 

When the ATS algorithm is applied to generate independent realisation for a
hypothesis test, one is able to construct a test for non-stationarity. If
two data sets do not fit the same distribution of ATS data then they can
not be said to be from the same deterministic dynamical
system. Unfortunately, the converse is not always true and the power of
the test depends on the choice of statistic. The utility of this
technique as a test for stationarity remains a subject for future investigation.

\section*{Acknowledgments}

This research was supported by a Hong Kong Polytechnic University
Research Grant (NO. A-PE46). The authors wish to thank J. Bhattacharya
for supplying the photo-plethysmographic time series.

\bibliographystyle{unsrt}

\end{document}